\documentclass[manuscript,IEEEtran]{article}
\usepackage{fullpage}

\usepackage{cite}
\usepackage{amsmath,amssymb,amsfonts}
\usepackage{xcolor}
\usepackage{float}
\usepackage{algorithm}
\usepackage{listings}
\usepackage{booktabs}
\usepackage{comment}
\usepackage{wrapfig}
\usepackage[colorinlistoftodos]{todonotes}
\def\BibTeX{{\rm B\kern-.05em{\sc i\kern-.025em b}\kern-.08em
    T\kern-.1667em\lower.7ex\hbox{E}\kern-.125emX}}
\usepackage{xspace}

\begin{document}

\title{
Reinventing High Performance Computing: Challenges and Opportunities
}

\author{
  Daniel Reed\\
  University of Utah\\
  \texttt{dan.reed@utah.edu}
  \and
  Dennis Gannon\\ 
  \texttt{dennis.gannon@outlook.com}
  \and
  Jack Dongarra\\
  University of Tennessee and Oak Ridge National Laboratory\\
  \texttt{dongarra@icl.utk.edu}
}







\maketitle
\begin{abstract}
%
The world of computing is in rapid transition, now dominated by a world of smartphones and cloud services, with profound implications for the future of advanced scientific computing.
Simply put,
high-performance computing (HPC) is at an important inflection point. For the last 60 years, the world's fastest supercomputers were almost exclusively produced in the United States on behalf of scientific research in the national laboratories.   
Change is now in the wind. 
While costs now stretch the limits of U.S. government funding for advanced computing,
Japan and China are now leaders in the bespoke HPC systems funded by government mandates.  
Meanwhile, the global semiconductor shortage and political battles surrounding fabrication facilities
affect everyone.
However, another, perhaps even deeper, fundamental change has occurred.   
The major cloud vendors have invested in global networks of massive scale systems that dwarf today's HPC systems.   
Driven by the computing demands of AI, these cloud systems are increasingly built using custom semiconductors, reducing the financial leverage of traditional computing vendors.
These cloud systems are now
breaking barriers in game playing and computer vision, 
reshaping how we think about the 
nature of scientific computation.
Building the next generation of leading edge HPC systems will require rethinking many fundamentals and
historical approaches by embracing end-to-end co-design; custom hardware configurations and packaging;
large-scale prototyping, as was common thirty years ago; and collaborative partnerships with the dominant computing ecosystem companies, smartphone and cloud computing vendors.
\end{abstract}

\section{Introduction}
\label{sec:intro}
Charles Dickens' classic, \textit{A Tale of Two Cities}, opens with this dichotomous line:
\begin{quote}
\textit{
IT was the best of times, it was the worst of times, it was the age of wisdom, it was the age of foolishness, it was the epoch of belief, it was the epoch of incredulity, it was the season of light, it was the season of darkness, it was the spring of hope, it was the winter of despair.
}
\end{quote}
It's an apt metaphor for the global state of information technology, and specifically for high-performance computing (HPC), with concomitant implications for how we adapt to changing technological constraints and opportunities.

Today, computing pervades all aspects of our society, in ways once imagined by only a few.  It now shapes how we socialize and communicate, engage in political discourse, purchase goods and services, conduct business and global commerce, explore new research ideas, and develop new technologies and products. 
Apple, Samsung, and Google now dominate the world of ubiquitous smartphones;
the large U.S. cloud and social media companies (i.e., Google, Microsoft, Amazon, and Facebook) dominate the U.S. NASDAQ with market capitalizations near or in excess of one trillion dollars, and their Asian counterparts (i.e., Baidu, Alibaba, and TenCent) are not far behind.

In turn, AI advances, coupled with large-scale computing infrastructure, have not only enabled highly accurate image and speech recognition \cite{trillionspeech}, quasi-autonomous vehicles and robots, they have led to scientific breakthroughs. As an example,
AlphaFold's recent success predicting 3-D protein structures \cite{alphafold} was
hailed as the 2021 {\em Science Breakthrough of the Year} \cite{breakthrough}.
More broadly, computing in all forms is an integral part of science and engineering, from laboratory instruments to global scientific facilities (e.g., LIGO \cite{ligo}, the Large Hadron Collider (LHC), the Vera Rubin Observatory \cite{rubin}, and the Square Kilometer Array \cite{ska}). 

Indeed, scientific computing has often been called the third paradigm, complementing theory and experiment, with big data and AI often called the fourth paradigm \cite{fourth-paradigm}.
Spanning both data analysis and disciplinary and multidisciplinary modeling, scientific computing systems have, like their commercial counterparts, grown ever larger and more complex, and today's exascale scientific computing systems rival global scientific facilities in cost and complexity.

However, not all is well, in the land of computing -- both commercial and scientific. Our society is now grappling with AI-generated deep fakes \cite{deepfake}, targeted political influencing \cite{echo}, privacy and security concerns, AI driven economic and social discrimination, global cybercrime, and threats from cyberwarfare.
Moreover, the global semiconductor shortage has highlighted the interdependence of global supply chains and the economic consequences for industries and countries, most dependent on a handful of global semiconductor foundries (e.g., Taiwan Semiconductor Manufacturing Company (TSMC), GlobalFoundries, and Intel).  

Concurrent with the rise of smartphones and cloud computing, the relative influence of governments and traditional computing companies on the scientific computing ecosystem has waned, with important implications for science and engineering discovery via leading edge high-performance computing (aka HPC or supercomputing).
Historically, government investments and the insights and ideas from designing and deploying supercomputers shaped the next generation of mainstream and consumer computing products. Today, that economic and technological influence has increasingly shifted to smartphone and cloud service companies. 
Moreover,  the end of Dennard scaling \cite{end-dennard}, slowdowns in Moore's Law, and the associated rising costs for continuing semiconductor advances, have made
building ever-faster supercomputers more economically challenging.

\begin{figure}[tb]
\centering
  \includegraphics[width=1.0\textwidth]{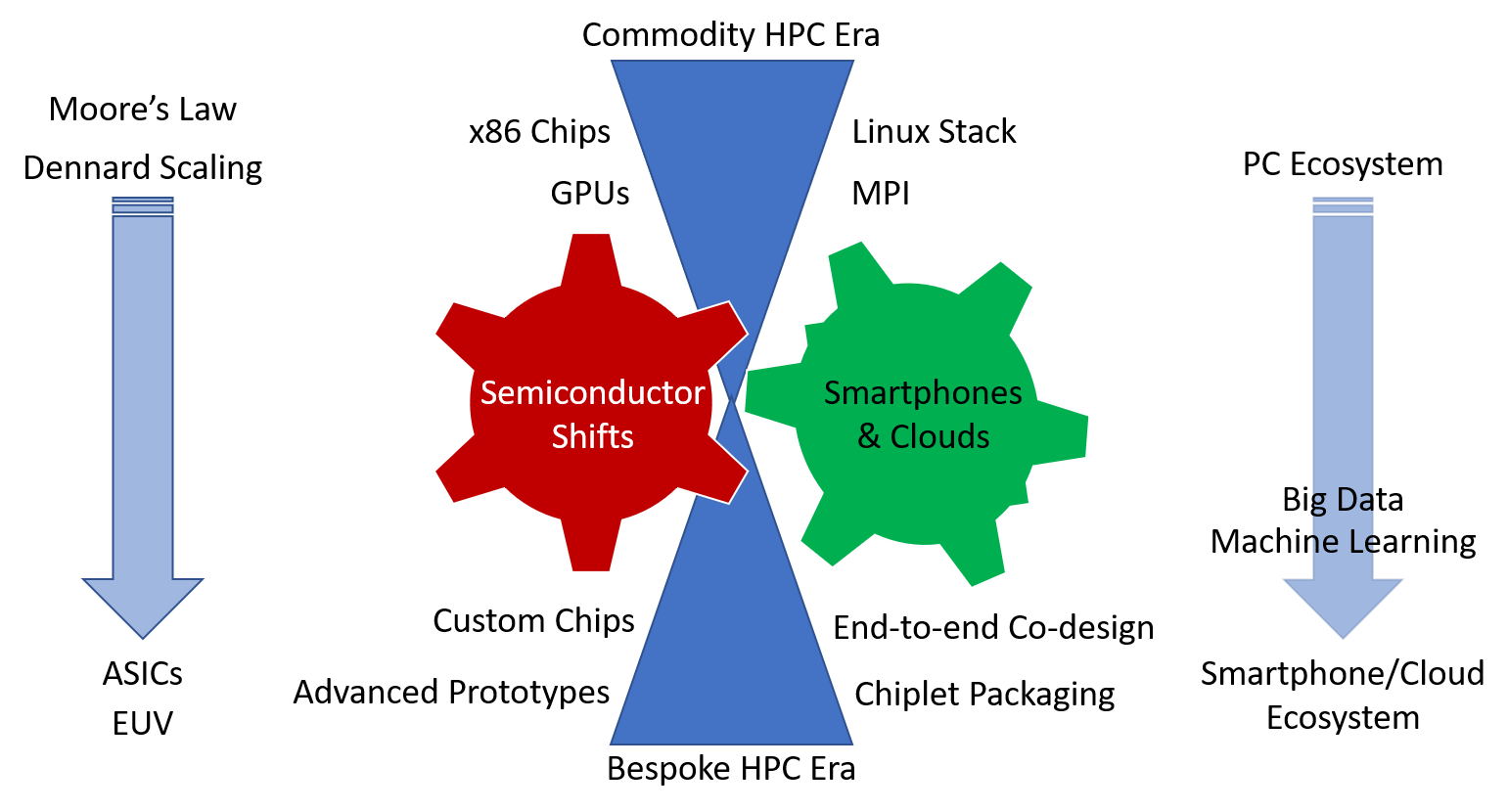}    
  \caption{Technical and Economic Forces Reshaping HPC}
  \label{fig:change}
\end{figure}
Given these seismic technology and industry shifts, the future of advanced scientific computing is now at a critical crossroads, particularly as the global competition for scientific computing leadership intensifies \cite{indicators}, raising several critical technological, economic, and cultural questions.
As Figure \ref{fig:change} suggests, our thesis is that current approaches
to designing and constructing leading edge high-performance computing systems must
change in deep and fundamental ways, embracing end-to-end co-design; custom
hardware configurations and packaging; large-scale
prototyping, as was common thirty years ago; and collaborative partnerships
with the dominant computing ecosystem companies, smartphone and cloud computing
vendors.

Let's begin by examining how all of this has happened, by following the money and the talent, then examining possible future directions for high-performance computing innovation and operations.
The remainder of this paper is organized as follows.  In \S\ref{sec:past} and \S\ref{sec:hpcpresent}, we briefly review the history of computing broadly and high-performance computing specifically. This is followed in \S\ref{sec:present} and \S\ref{sec:shifts} by a discussion of the dramatic economic and technological shifts occurring in cloud computing. In turn, \S\ref{sec:chiplets} discusses the semiconductor issues that are driving the adoption of multichip modules (aka chiplets).
Finally, \S\ref{sec:future} summarizes the current technological and economic state, and
\S\ref{sec:hpcfutures} assesses potential future directions for leading edge scientific computing.
\section{A Computing Precis}
\label{sec:past}
To understand the present and the possible future of HPC, it is instructive to examine the past, with each layer of the computing archaeological dig -- mainframes, minicomputers, workstations, personal computers, and mobile devices revealing nuance but also repeated lessons.
In that spirit, consider a {\em gedanken} experiment, where one asks a group of people to name the first computer company that comes to mind.

Older individuals might mention IBM, Amdahl, and the BUNCH (Burroughs, UNIVAC, NCR, CDC, and Honeywell), the erstwhile mainframe competitors with IBM.
Although IBM once bestrode the computing world like a colossus, today, even its relevance as faded, as newer competitors have grown and flourished. Indeed, few outside computing circles remember that IBM once made a "bet the company'' investment in the System/360 \cite{s360} line of instruction set compatible systems that spanned a wide range of performance and set the standard for business data processing.
Nor will most remember that IBM was once locked in a bitter struggle with the U.S. Department of Justice for monopolistic practices under anti-trust laws.  

Somewhat younger individuals might mention Data General's Eclipse and Digital Equipment Corporation's (DEC)  VAX minicomputers, while others would speak wistfully about SUN and Apollo workstations or Symbolics or Xerox Lisp machines. Those even younger would speak nostalgically about the Apple Macintosh and the IBM PC and its clones, with those born in the mobile era discussed Apple's iPhone, Google's Pixel, or Samsung's Galaxy smartphones.

As for the BUNCH, they have either disappeared from the computer business or are niche players at best. Later, many others, including several market leaders – Compaq, Tandem, Data General, DEC, SUN, Apollo, Symbolics, and Silicon Graphics (SGI) –- came and went.

\begin{figure}[tb]
\centering
  \includegraphics[width=1.0\textwidth]{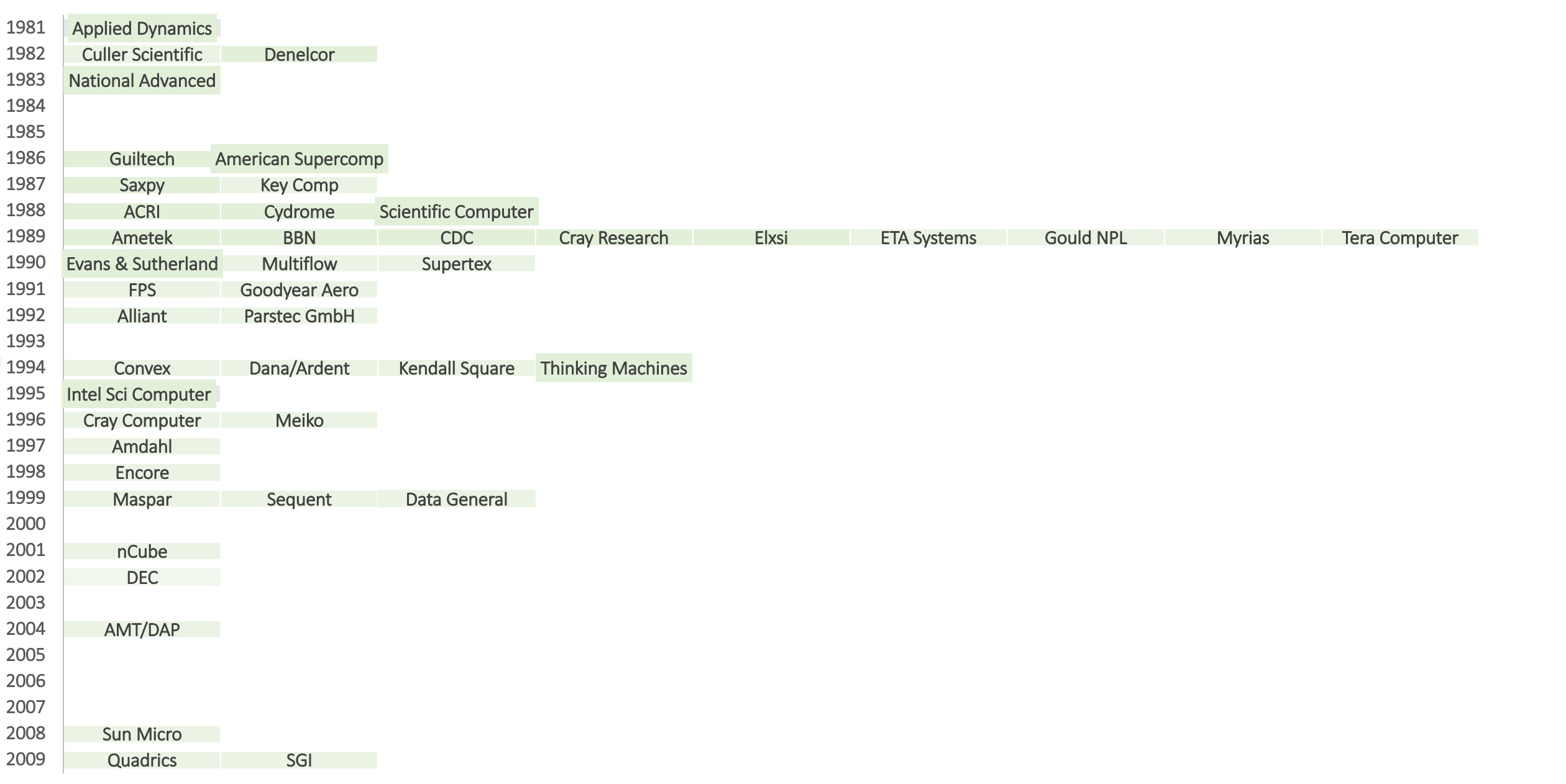}    
  \caption{The Rise and Fall of Commercial HPC Companies (Date of Market Exit)}
  \label{fig:deadcompanies}
\end{figure}

There's a similar tale in the land of high-performance computing, where the technology cemetery is filled with departed denizens, some large, some small, but all now permanent members of the Dead Computer Society \cite{dcs}.
Notable high-end examples include ETA Systems, Supercomputer Systems (SSI), Tera Computer, Kendall Square Research (KSR), and Thinking Machines, as well as minisupercomputer companies such as Alliant and Convex.
Whether by misjudging markets or technologies, each in its own way proved the truth of the old adage that the best way to make a small fortune in supercomputing is by starting with a large fortune, then shipping one or two generations of product, as Figure \ref{fig:deadcompanies} shows.

As a recent example of the Darwinian struggle for financial stability in high-performance computing, SGI first acquired Cray Research, the company led by Seymour Cray. Before its own demise, SGI sold the Cray business to Tera Computer, which resurrected the Cray name, only to be acquired by HPE.
This leaves HPE as one of the few high-end integrators of leading edge high-performance computing (HPC) systems in the United States.  

As these examples illustrate, computing system technologies and economics shift rapidly.
The history of computing is replete with examples of seemingly promising technologies felled by large development costs and relatively small markets.
This is particularly true in leading edge computing of all types.

\section{High-Performance Computing: Past and Present}
\label{sec:hpcpresent}
\begin{figure}[tb]
\centering
  \includegraphics[width=1.0\textwidth]{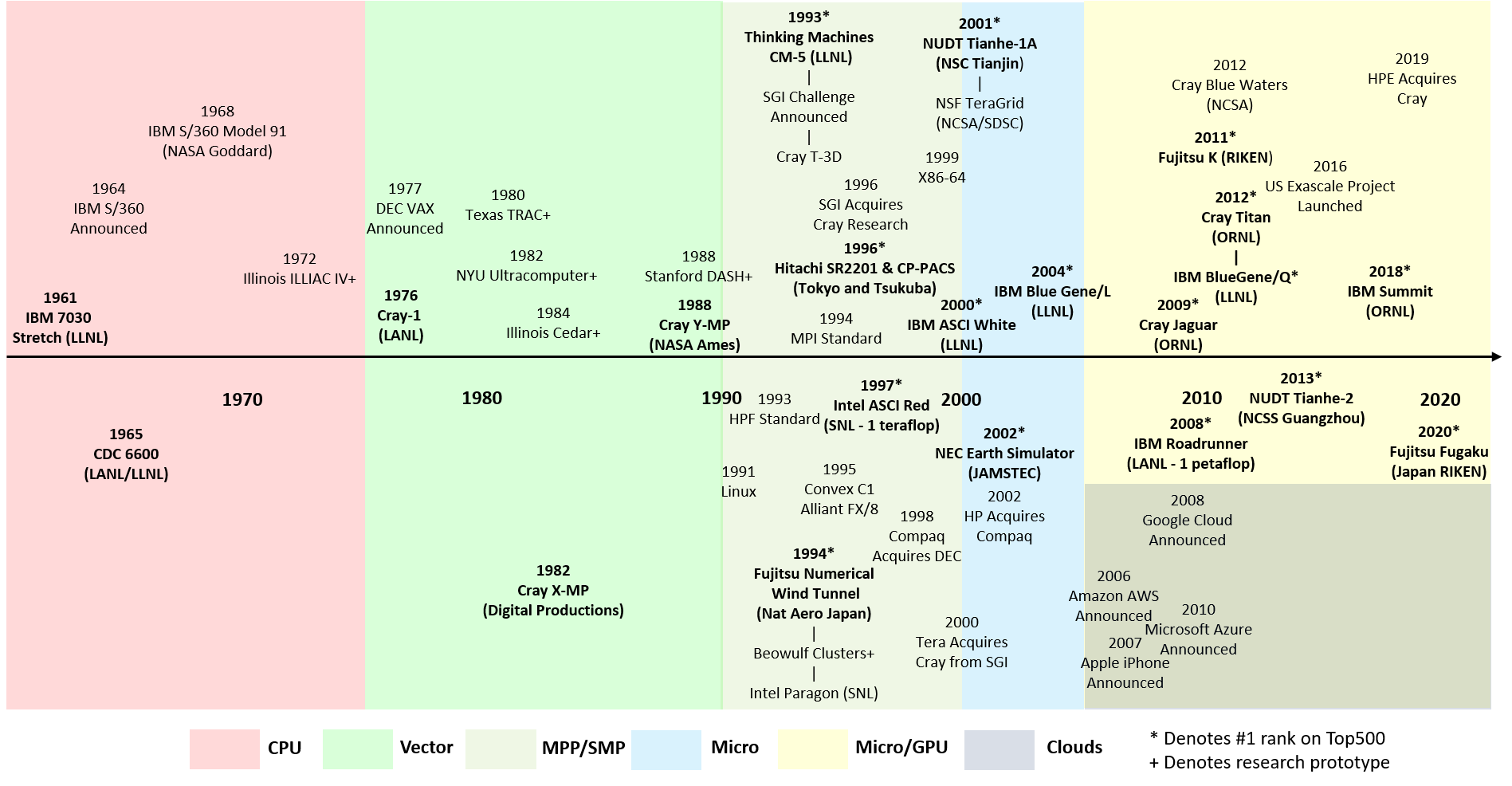}    
  \caption{Timeline of Advanced Computing}
  \label{fig:timeline}
\end{figure}

In the nearly fifty years since the introduction of the Cray-1 \cite{cray-1} in 1975, 
the HPC market has seen many changes. 
Figure \ref{fig:timeline} illustrates some of the key events in the history of high-performance computing, along with recent systems ranked as the fastest in the world.
Initially, the HPC market targeted systems clearly distinct from the mainstream, with a wide range of academic research, prototyping, and construction projects (e.g., ILLIAC IV \cite{illiac4}, Cosmic Cube \cite{cosmiccube}, WARP \cite{warp}, Cedar \cite{cedar}, Ultracomputer \cite{ultracomputer}, DASH \cite{dash}), each with
specialized processors and other hardware, custom operating systems, and optimizing compilers. 

Today, the HPC market is no longer an isolated niche market for specialized systems. Vertically integrated companies produce systems of any size, and the components used for these systems are the same, whether for an individual server or the the most powerful supercomputers. Similar software environments are available on all of these systems, and as noted earlier, market and cost pressures have driven most users and centers away from specialized, highly integrated  supercomputers toward clustered systems built using commodity components.

To provide a reliable basis for tracking and detecting trends in high performance computing, the TOP500 project \cite{top500} was started in 1993. The TOP500 list (www.top500.org) has served as the defining yardstick for supercomputing performance since then. 
Published twice a year, it compiles the world's 500 largest computers and some of their main characteristics. Systems are ranked according to their performance of the Linpack benchmark \cite{linpackppf}, which solves a dense system of linear equations. Over time, the data collected for the list has enabled the early identification and quantification of many significant technological and architectural trends related to high-performance computing.

In the second half of the 1970s, the introduction of vector computer systems, as exemplified by the Cray-1, marked the beginning of modern supercomputing. These systems offered a performance advantage of at least one order of magnitude over conventional systems available at that time, with raw hardware performance was the main if not the only selling argument. 

In the first half of the 1980s, the integration of vector systems in conventional computing environments became more critical in U.S. markets.
The surviving manufacturers began providing standard programming environments, widely used operating systems (e.g, variants of Unix), and critical applications, allowing them to attract customers from the broader industrial market. 
Performance was mainly increased by improved chip technologies and by producing shared-memory multiprocessor (SMP) systems.

During this time, the commercial viability of vector processors and high-bandwidth memory subsystems in the U.S., (e.g., from Cray and ETA) and in Japan (e.g, from Fujitsu, NEC, and Hitachi) became increasingly problematic, given their high costs.
However, at least one large company in Japan (NEC) remains committed to traditional vector architectures.

Fostered by several U.S. government programs, massively parallel computing  using distributed memory became the center of interest at the end of the 1980s. Stimulated by the Caltech Cosmic Cube \cite{cosmiccube}, the Intel Paragon XP/S \cite{paragon}, nCube10 \cite{ncube} and Thinking Machines CM-5 \cite{cm5} overcame the hardware scalability limitations of shared memory systems. However, the increased performance of standard microprocessors after the reduced instruction set
computing (RISC) revolution \cite{patterson}, together with the cost advantage of volume microprocessor production, soon formed the basis for the ``Attack of the Killer Micros'' \cite{brooks,nyt} and
led to the demise of most bespoke HPC systems.  

In the early 2000s, Beowulf clusters \cite{beowulf} built with off-the-shelf components gained prominence as academic research objects and as computing platforms for end-users of HPC computing systems. By 2004, these clusters, with their relatively low cost and comparatively high performance, represented the majority of new systems on the TOP500 in a broad range of application areas.
In 2001, for example, the NSF-funded National Center for Supercomputing Applications (NCSA) deployed two one teraflop Linux clusters, quickly followed by the even larger NSF TeraGrid \cite{teragrid}. 
\begin{figure}[tb]
\centering
  \includegraphics[width=1.0\textwidth]{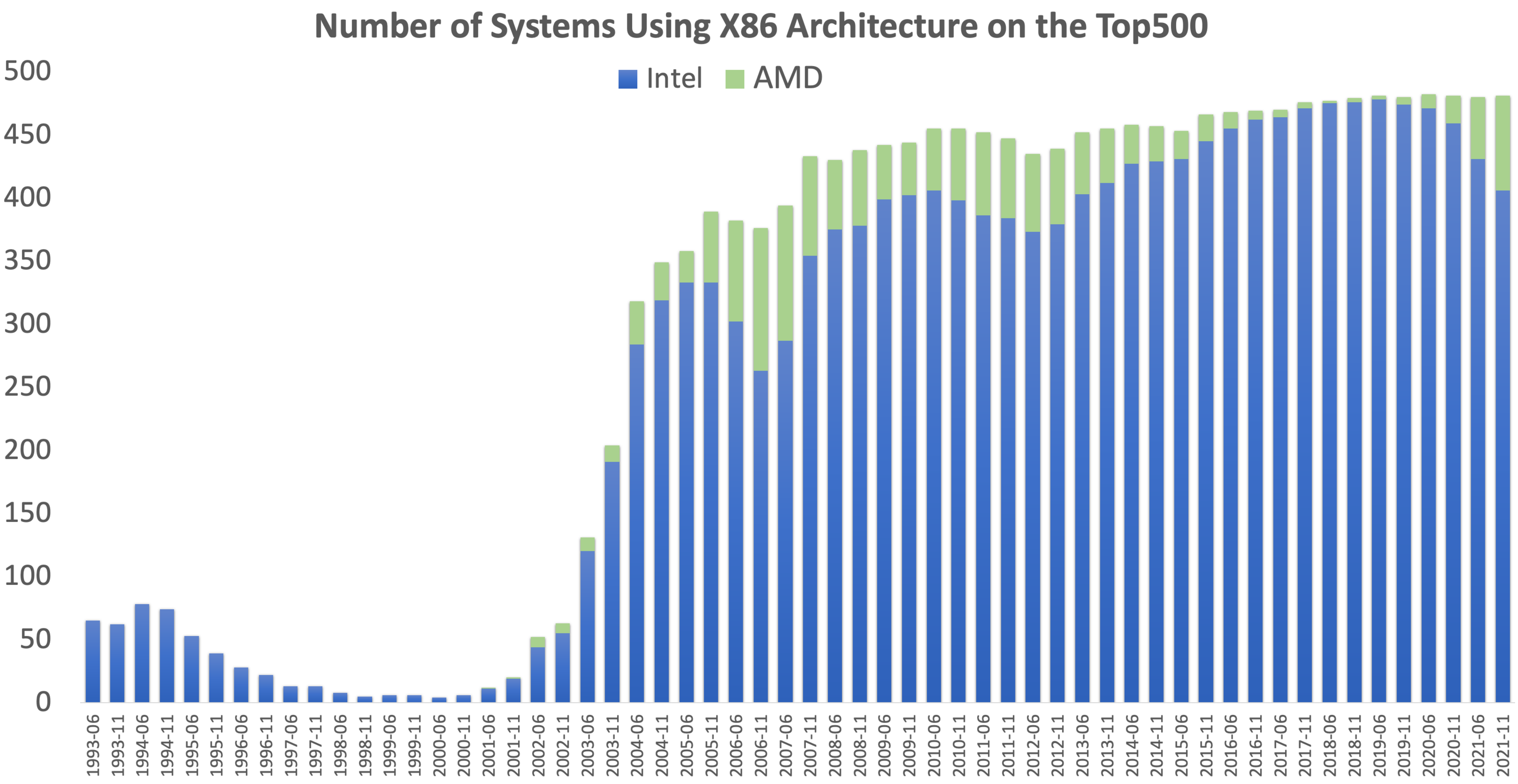}    
  \caption{Systems Using the x86-64 Architecture on the TOP500}
  \label{fig:chartx}
\end{figure}

Since 2005, graphics processing units (GPUs), predominantly developed by NVidia, have emerged as a powerful platform for high-performance computation and have been successfully used to accelerate many scientific workloads. Typically, the computationally intensive parts of the application are offloaded to the GPU, which serves as the CPU's parallel co-processor. Today's modern supercomputer systems, in practice, consist of many nodes, each holding between 2 and 32 conventional CPUs and 1–6 GPUs. There will usually also be a high-speed network and a system for data storage. 
%
\begin{figure}[tb]
\centering
   \includegraphics[width=1.0\textwidth]{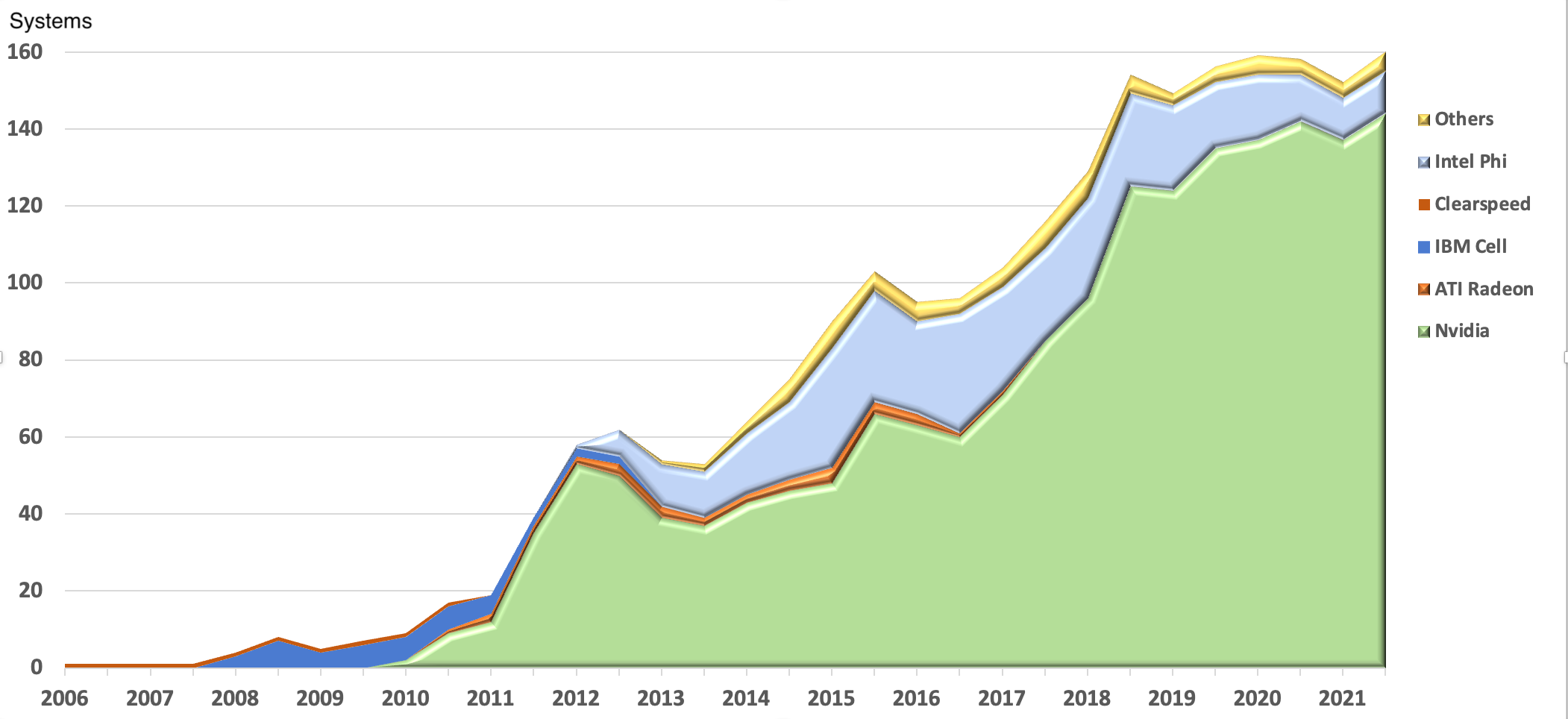}
   \caption{Systems Using GPU Accelerators on the TOP500}
   \label{fig:gpus}
\end{figure}

When the TOP500 list began in 1993, only 65 systems  used Intel's i860 architecture \cite{i860}, and the rest had specialized architectures. The most recent TOP500 list shows 81\% of the systems are based on Intel processors, and another 15\% are based on AMD processors (i.e., 96\% of the systems on the list are based on the x86-64 architecture and many use GPU accelerators); see Figures \ref{fig:gpus} and \ref{fig:chart1}. 
In 1993, the  systems on the TOP500 list used a variety of custom operating systems; today, all 500 systems run Linux. The situation with interconnects is similar, with 240 machines using gigabit Ethernet and 180 systems using InfiniBand \cite{infiniband}.

In brief, the systems on the TOP500 today are largely examples of a commodity monoculture, built from nodes containing  server-class microprocessors and the same GPU accelerators widely used for machine learning.  In turn, they are connected by either high-speed Ethernet or InfiniBand and run some variant of Linux.
However, despite these large technological similarities, there are global differences, driven by a combination of national politics, economics, and national priorities.

In Japan and the U.S., the issues and concerns about HPC have proven broadly similar.
HPC continues to be critical for many scientific and engineering pursuits, including climate modeling, earthquake simulation, and biosystems. However, Japan lacks the kind of defense missions,
such as nuclear stockpile stewardship, that have historically been drivers for U.S.
supercomputing. As a result, investment in Japanese supercomputers must be justified based on 
their ultimate economic and societal benefits for a civil society.

The launch of the Japanese NEC Earth Simulator (ES) \cite{earthsim} in 2002 created a substantial 
concern in the United States that the U.S. had lost its lead in high performance computing. 
While there was undoubtedly a loss of national pride
because a supercomputer in the United States was not first on a list of the
world's fastest supercomputers, it is important to understand the set of
issues surrounding the loss of that position. The development of the ES
required a significant Japanese investment (approximately \$500 million, including the
cost of a special facility to house the system) and a commitment over a
long period. The United States made an even more significant investment in
HPC under the Accelerated Strategic Computing (ASC) program, but the money was not spent on a single
platform, rather it funded several systems over multiple years, beginning with the ASCI Red terascale cluster at Sandia National Laboratory \cite{ascired}. 
Other significant differences between the Japanese and U.S. systems include the following:
\begin{itemize}
\item The ES was developed for basic research and access was shared internationally, whereas the ASC program was driven by national defense and its systems have been used only for domestic missions.
\item A large part of the ES investment supported NEC’s development of its SX-6 technology. 
In contrast, the ASC program has made only modest investments in industrial R\&D.
\item The ES used custom vector processors, whereas ASC systems largely used commodity microprocessors.
\item The ES software technology primarily came from abroad, although it was often modified and enhanced in Japan. For example, many ES codes were developed using a Japanese-enhanced version of High Performance Fortran \cite{hpf}. Virtually all software used in the ASC program has been produced in the United States.
\end{itemize}
Most recently, Fujitsu's development of the Fugaku supercomputer, based on custom processors with ARM cores and vector instructions \cite{fugaku}, has followed in this Japanese tradition.

Although Europe has launched several advanced computing projects, it remains largely dependent on U.S. vendors for advanced computing technology.
Collectively, the European Union countries hosted 48 of the TOP500
systems as of November 2021; this amounts to 9.6\% percent of the TOP500 listed
systems and 8.6\% percent of the Top500's total computing capability. However, it is not
clear that one should treat the European Union as a single entity. 

Perhaps the major exception to dependence on U.S. technologies has been China.
China's high performance computing centers have developed their specialized processors and interconnect hardware due to U.S. export controls. As a result, the export controls have only strengthened China's determination to invent its own technology \cite{china}.


In the United States, the Exascale Computing Initiative (ECI) was launched in
2016 by the Department of Energy’s (DOE) Office of Science and the National
Nuclear Security Administration, as a research, development, and deployment
project focused on the delivery of mission-critical applications, an
integrated software stack, and exascale hardware technology advances.
The Exascale Computing Project \cite{ecp} (ECP), a component of the ECP, focused on a range of applications, including
chemistry and materials; energy production and transmission; earth and space science; data analytics and optimization; and national security, along
with application frameworks to support application portability and efficiency.


The total cost of the ECP effort is projected to be \$1.8B over the seven years of the project. The ECP budget does not include the procurement of the first three exascale systems (Frontier (ORNL), Aurora (ANL), and El Capitan (LLNL)). Each of those exascale systems costs roughly \$600M. Which brings the cost of ECP plus hardware procurement to around \$3.6B.

\section{Shifting Scale and Ecosystem Influence}
\label{sec:present}
Returning to our {\em gedanken} experiment, for most computing system users, particularly those classified as
Gen Z or Gen Alpha, their view of computing is defined by desktop computers, Apple, Google, and Samsung smartphones, and the cloud services that support them.
Similarly, the rapid growth of commercial cloud services and business outsourcing has made Amazon AWS, Microsoft Azure, and Google Cloud among the fastest growing elements of the computing services industry.
Put another way, the FAANG companies (Meta née Facebook, Amazon, Apple, Netflix, and Alphabet née Google) plus Microsoft and their Asian BAT (Baidu, Alibaba, and Tencent) counterparts are of a financial scale perhaps unprecedented since the early 20th century days of the Standard Oil monopoly.  

\begin{figure}[tb]
\centering
  \includegraphics[width=1.0\textwidth]{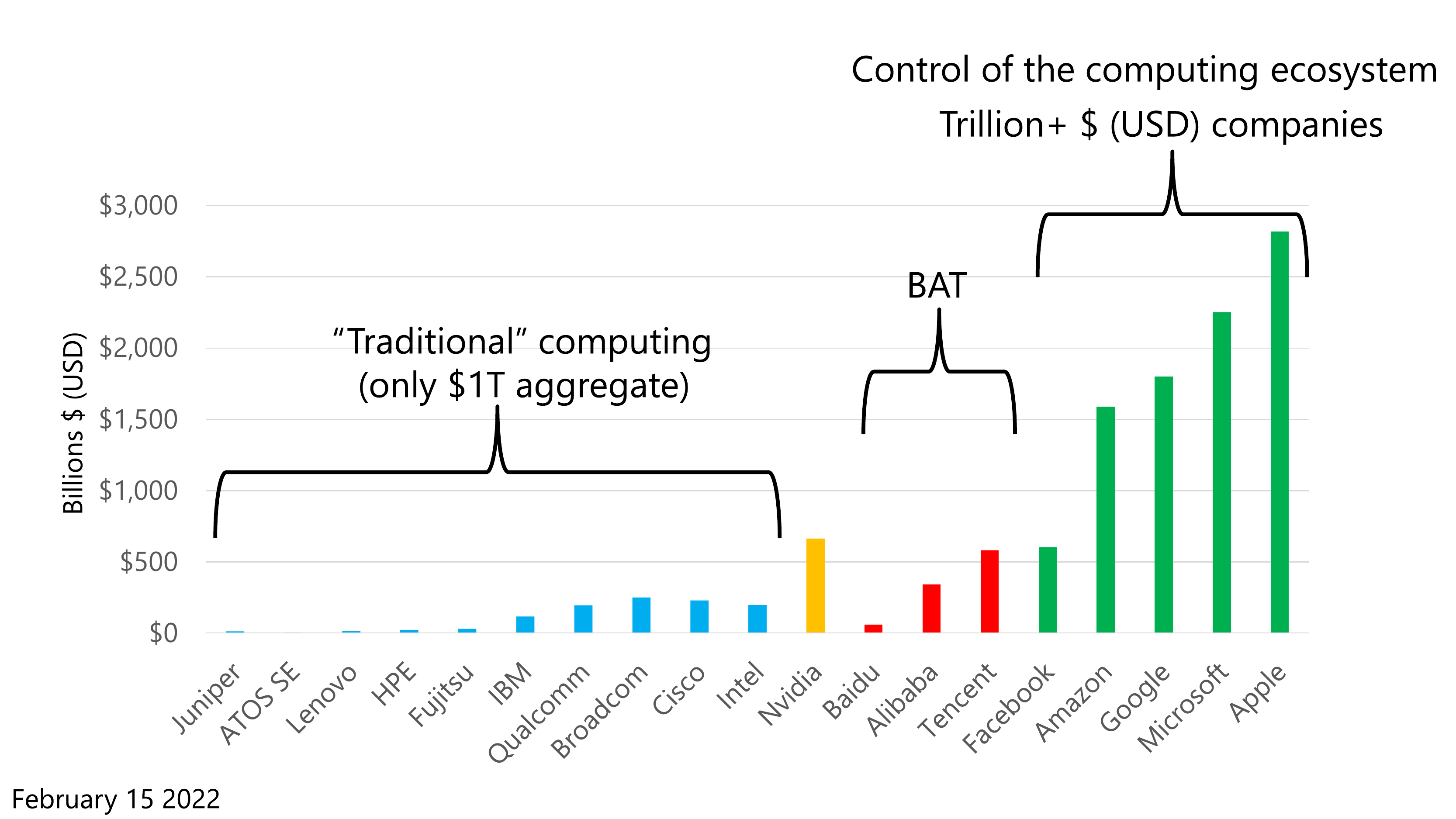}                   
  \caption{Computing Company Market Capitalization}
  \label{fig:chart1}
\end{figure}

%
As  Figure \ref{fig:chart1} shows, the market capitalizations of the smartphone, social media, and cloud companies now dwarf that of those companies that manufacture microprocessors, networking equipment, and other computing components. 
NVidia is the notable exception, fueled by the growing demand for GPUs to support both computer gaming and AI.
These ecosystem shifts have profound implications for the future of high-performance computing, including which groups drive both research and development and system configuration and access, a topic to which we will return in \S\ref{sec:future}.
To be sure, market capitalization is not the only metric of influence.  However, is an indicator of financial leverage, albeit one conditioned in part on stock market assessments.

Just as the PC revolution displaced incumbents in the mainframe, minicomputer, and workstation markets, smartphones and cloud services companies are increasingly displacing the PC market incumbents.
Throughout the PC era, the Microsoft and x86 microprocessor combination (the Wintel duopoly) locked Microsoft and Intel, and to a lesser extent, AMD, in a collaborative/competitive partnership as rough peers.  
As server chips, clouds, and infrastructure grew in importance relative to desktops and laptops, that power
dynamic began to shift.  
Today, Intel and AMD must be responsive to cloud vendor requirements, as a large fraction their total
microprocessor production is purchased by those vendors, as their quarterly financial statements demonstrate.
Put another way, quantitative changes in scale and capabilities have catalyzed qualitative changes in capabilities and influence.

Economies of scale first fueled commodity HPC clusters and attracted the interest of vendors as large-scale demonstrations of leading edge technology.
Today, the even larger economies of scale of cloud computing vendors has diminished the influence of high-performance computing on future chip and system designs.  
No longer do chip vendors look to HPC deployments of large clusters as flagship technology 
demonstrations that will drive larger market uptake.
\section{The Rise of Cloud Services}
\label{sec:shifts}
To understand the potential future of high performance computing, one must examine the fundamental shifts in computing technology. These shifts have occurred along two axes:
the rise of massive scale commercial clouds and the economic and technological challenges associated with
the evolution of semiconductor technology.  

As we noted in \S\ref{sec:present}, Apple, Samsung, Google, and the other cloud service companies now dominate the computing hardware and software ecosystem, both in scale and in technical approaches. Initially, these companies purchased standard servers and networking equipment for deployment in traditional collocation centers (colos). As scale increased, they began designing purpose-built data centers, optimized for power usage effectiveness (PUE) \cite{PUE}, deployed at sites selected via multifactor optimization – inexpensive energy availability, tax incentives and political subsidies, political and geological stability, network access, and customer demand.

The early cloud data centers were built to support specific tasks such as email and web search. While they were large for their time, they are now dwarfed by the scale of current deployments.  The driving force behind this explosive growth was the realization that the data center itself was a consumer product.  In 2006, Amazon launched its "Simple Storage Service" (S3), allowing anyone with a credit card to store data ''in the cloud.''  A few months later, Amazon introduced the "Elastic Compute Cloud" (EC2) built upon virtual machine technology. This proved a boon for start-up companies who needed an on-line presence but lacked the resources to build a data center.  EC2 could automatically expand as a business grew.  

Amazon Web Services (AWS), as the new division of Amazon was called, made S3 and EC2 more valuable by releasing relational database services and DynamoDB, a "noSQL" on-line database service. By 2008, Microsoft was developing its own counterpart to AWS, with Microsoft Azure announced in 2008 and released in 2010 with basic cloud storage and virtual machine capabilities.   While Google already had a very large data center deployment to support search and advertizing, as well as and numerous internally developed cloud services, its  first public services, AppEngine was released in 2011 \cite{appengine}.  

An explosion in cloud capabilities followed these early services. In addition,  the growing movement of open source software began driving dramatic changes in the industry. Open source cloud platforms such 
as Eucalyptus \cite{eucalyptus} and OpenStack \cite{openstack} allowed others to start building their own cloud platforms. Microservices  then emerged, which enabled the partitioning of massive-scale, multiuser cloud applications into many small interdependent service units. In turn, this allowed the applications to scale and be upgraded while running uninterrupted.  

Ecosystem ferment continued with Google's open source release of its Kubernetes microservice framework \cite{kubernetes} in 2014. 
Kubernetes is a container orchestration service that supports software packaged with tools like the open-source Docker container system \cite{docker}.  Together Kubernetes, Docker, and a few other systems form the foundation of the ''Cloud Native Software'' movement, which is now supported by all major cloud vendors. (Versions of Kubernetes and container services are now supported on several supercomputer platforms \cite{sdsc} as well.) 
Serverless computing has proven to be another important addition to the software stack.  This allows computation to be fully abstracted from the underlying infrastructure, allowing computation to migrate between the cloud and edge computing devices. 

As cloud scale, complexity, and operational experience continued to grow, additional optimization and leverage opportunities emerged, including software defined networking \cite{sdn}, protocol offloads, and custom network architectures (greatly reducing dependence on traditional network hardware vendors) \cite{smartnic}; quantitative analysis of processor \cite{core}, memory \cite{dram1,dram2}, network \cite{net1,net2} and disk failure modes \cite{disk1,disk2}, with consequent redesign for reliability and lower cost (dictating specifications to vendors via consortia like Open Compute \cite{opencompute}); custom processor SKUs, custom accelerators (FPGAs and ASICs), and finally, complete processor design (e.g., Apple silicon, Google TPUs \cite{tpu} and AWS Gravitons).  In between, the cloud vendors
deployed their own global fiber networks.

While cloud native software frameworks have become standard tools for developers, they are not the top of the software stack.  Enterprise customers were most interested in services to better serve customers. 
As mentioned above, for cloud companies, the differentiating services has become machine learning and other AI capabilities such as voice-to-text and text-to-speech, document analysis, fraud detection, recommendation services, image and video analysis, and demand forecasting.  
Amazon released its SageMaker AI toolkit \cite{sagemaker} and a host of ready to use AI tools, Google Cloud supports a variety of cloud AI services backed  by Tensor Processing Units (TPUs) \cite{tpu}, and  Microsoft Azure introduced Cognitive Service and their ML workspaces development environment. 
Tensorflow \cite{tensorflow} from Google and Torch from Meta have become the open source industry standard tools for AI development.  

Nor has the cloud push for AI stopped there. All of cloud vendors began buying or partnering with the most advanced AI startups.  Google acquired DeepMind in the U.K. in 2014, Microsoft partnered with OpenAI in 2019, and Meta (Facebook) and Amazon have each acquired several smaller AI startups. Both DeepMind and OpenAI have been using the infrastructure of their parent/partner cloud companies to build massive semi-supervised language models that are changing the game for AI.
%

What is the distinguishing value proposition that has allowed cloud service vendors to grow so large so rapidly? For small companies, the ability to launch backend and customer facing software services without the need to first purchase, deploy, and support on-premise hardware both reduces initial capital requirements and the time to market. Talent and expertise can focus on the key business proposition and the associated services can scale quickly in response to rising customer demand.  

For large companies, those same scaling properties apply, along with a focus on the differentiating properties of their business rather than IT staff and infrastructure.  In both cases, companies benefit from cloud services geo-distribution, with associated business continuity risk reduction, cybersecurity protections, inter-company data sharing via standardized services and protocols, regular hardware upgrades, and a rapidly expanding set of software services and features.

This virtuous cycle of insatiable consumer demand for rich services, business outsourcing to the cloud, expanding data center capacity, and infrastructure cost optimization has had several effects.  First, it has dramatically lessened – and in many cases totally eliminated -- FAANG/Microsoft/BAT dependence on traditional computing vendors and ameliorated the risks of wholesale transfer pricing. Put another way, Amazon AWS, when necessary, can now negotiate pricing terms with Intel and AMD in the same way Walmart does with Proctor and Gamble (i.e., from a position of strength as a large customer).

Second, the absolute scale of infrastructure investment, denominated in billions of dollars per year for each cloud service vendor, means these companies are shaping computing technology research and development priorities in ways traditional computing vendors cannot, driving design and fabrication priorities in every element of the ecosystem. Bigger is not simply bigger; bigger is different, fueling investment in new technologies at a phenomenal rate.  Relatedly, the massive scale of these deployments far exceeds anything seen in business, academia or government, with the already large gap widening each year.  Today, the compound annual growth rate (CAGR) for cloud services is 50 percent at multiple companies.

Third, custom infrastructure is about more than economic leverage; it's about market differentiation and competitive advantage.  Anything not differentiating is standardized and commoditized, driving unnecessary cost out of the system, and concomitantly reducing profits for commodity purveyors.  Conversely, anything differentiating is the focus of intense internal optimization and aggressive deployment.

Equally importantly, client plus cloud services vendors predominately make their money from the services offered atop their infrastructure, much less from the infrastructure itself. Unlike Intel, AMD, and Nvidia, which sell silicon, Apple is not selling its A15 SoCs; it is selling iPhones, just as Amazon and Google are selling services atop TPUs and Gravitons.

\section{Semiconductor Shifts}
\label{sec:chiplets}
For the past twenty years, the most reliable engine of performance gains has been the steady
rhythm of semiconductor advances -- smaller, faster transistors and larger, higher performance chips. 
However, as chip feature sizes have approached 5 nanometers and Dennard scaling ended \cite{end-dennard}, the cadence of new technology generations has slowed, even as
semiconductor foundry costs have continued to rise.
With the slowing of Moore's Law and
the end of Dennard scaling \cite{end-dennard}, improved performance of supercomputers has increasingly
relied on larger scale (i.e., building systems with more computing elements) and GPU co-processing. 
However, this has not lessened the desire to build higher performance processors.

With the shift to extreme ultraviolet (EUV) lithography \cite{euv} and gate-all-around FETs \cite{gaa},
the ''minimax problem'' of maximizing chip yields, minimizing manufacturing costs, and maximizing chip
performance has grown increasingly complex.
Chiplets \cite{chiplet1, chiplet2} have emerged as a way to address these issues, while also 
integrating multiple functions in a single package.
The chiplet packaging concept is old and simple. 
Rather than fabricating a monolithic system-on-a-chip (SoC), chiplet technology combines
multiple chips, each representing a portion of the desired functionality, possibly fabricated
using different processes by different vendors and perhaps including IP from multiple sources.

Chiplet designs are part of the most recent offerings from Intel and AMD, where the latter's EPYC and 
Ryzen processors have delivered industry-leading performance via chiplet integration \cite{chiplet2}.
Similarly, Amazon's Graviton3 also uses a chiplet design with seven different chip dies, and its
Advanced Query Accelerator (AQUA) for AWS Redshift, Amazon's powerful and popular data warehouse service,
relies on a package of custom ASICs and FPGA accelerators. 


\section{A Computing Checkpoint}
\label{sec:future}
As we have described, the world of computing is in flux, driven by economic, technological, and political realities:  
dark silicon, functional specialization, chiplets, semiconductor fabrication challenges, rising
fabrication facility costs, and global economic and political competition for foundry self-reliance.
Assembling the technology and economics puzzle pieces yields the following truisms:
\begin{itemize}
\item	Advanced computing of all kinds, including high-performance computing, requires ongoing non-recurring engineering (NRE) investment to develop new technologies and systems.  
\item	The smartphone and cloud services companies are cash rich (i.e., exothermic), and they are designing, building, and deploying their own hardware and software infrastructure at unprecedented scale.
\item   The software and services developed in the cloud world are rich, diverse, and rapidly expanding, though only some of them are used by the traditional high-performance computing community.
\item	The traditional computing vendors are now relatively small economic players in the computing ecosystem, and many are dependent on government investment (i.e., endothermic) for the NRE needed to advance the bleeding edge of advanced computing technologies.
\item	AI is fueling a revolution in how businesses and researchers think about problems and their computational solution.
\item	Dennard scaling \cite{end-dennard} has ended and continued performance advances increasingly depend on functional specialization via custom ASICs and chiplet-integrated packages.
\item	Moore's Law is at or near an end, and transistor costs are likely to increase as features sizes
continue to decrease.
\item Nimble hardware startups are exploring new ideas, driven by the AI frenzy.
\item	Talent is following the money and the opportunities, which are increasingly in a small number of very large companies or creative startups.
\end{itemize}
These changes make it obvious that a computing revolution is underway, one that is reshaping global
leadership in critical areas.  All revolutions create opportunity, but only for those prepared to accept
the risks and act accordingly. 

With this backdrop, what is the future of advanced computing? Some of it is obvious, given the current dominance of cellphone vendors and cloud service providers. 
However, it seems likely that continued innovation in advanced high-performance computing  will require rethinking some of our traditional approaches and assumptions, including how, where, and when academia, government laboratories, and companies spends finite resources and how and we expand the global talent base.  
Let's start with the broad issues around the the information technology ecosystem, viewed from a national, regional, and global policy perspective.

First, there are substantial social, political, economic, and national security risks for any country that lacks a robust silicon fabrication ecosystem.
Fabless semiconductor firms are important, but onshore, state of the art fabrication facilities are critical, as the ongoing global semiconductor shortage has shown. However,  the investment needed to build state of the art facilities is denominated in billions of dollars per facility.  
In the U.S., the the U.S. CHIPS Act and its successors, which would provide government support, are topics of intense political debate, with similar conversations underway in the European Union.
Finally, Intel, TSMC, and GlobalFoundries recently announced plans to build new chip fabrication facilities in the U.S., each for different reasons.

Second, the seismic business and technology shifts have also led to equally dramatic talent shifts. People gravitate to opportunities, many students, faculty, and staff are leaving academia, national laboratories, and traditional computing companies to pursue those opportunities.  The opportunity to be creative will always be a draw for the most talented.  Seymour Cray once said, “I wanted to make a contribution,” when asked why he had used so many high risk technologies in the Cray-1. That
desire is no less true today.

The academic brain drain among artificial intelligence researchers is well documented.  After all, in the commercial world one can now develop and test ideas at a scale simply not possible in academia, using truly big data and uniquely scaled hardware and software infrastructure.  The same is true for chip designers and system software developers.  
These talent challenges are important national policy considerations for all industrialized countries.

With this backdrop, what are the implications for high-performance computing futures?

\section{Leading Edge HPC Futures}
\label{sec:hpcfutures}
It now seems self-evident that supercomputing, at least at the highest levels, is endothermic, requiring regular infusions of non-revenue capital to fund the non-recurring engineering (NRE) costs to develop and deploy new technologies and successive generations of integrated systems. In turn, that capital can come from either other, more profitable divisions of a business or from external sources (e.g., government investment).  

Cloud service companies now offer a variety of high-performance computing clusters, of varying size, performance, and price.
Given this, one might ask why cloud service companies are not investing even more deeply in the HPC market?
Any business leader must always look at the opportunity cost (i.e., the time constant, the talent commitment, and cost of money) for any NRE investments and the expected return on those investments.  
The core business question is always how to make the most money with the money one has, absent some other marketing or cultural reason to spend money on loss leaders, bragging rights, or political positioning.   
The key phrase here is ''the most money;'' simply being profitable is not enough.  

Therein lies the historical ''small fortune'' problem for HPC.  The NRE costs
for leading edge supercomputing are now quite large relative to the revenues
and market capitalization of those entities we call ``computer companies,''
and they are increasingly out of reach for most government agencies, at least under
current funding envelopes. The days are long past when a few million dollars could buy a
Cray-1/X-MP/Y-MP/2 or a commodity cluster and the resulting system would land
the top ten of the TOP500 list.
Today, hundreds of millions of dollars are needed to deploy a machine near the
top of the TOP500 list, and at least similar, if not larger, investments in
NRE are needed.

What does this brave new world mean for leading edge HPC?
We believe {\bf six maxims} must guide future HPC government and private
sector research and development strategies, for all countries.

{\bf Maxim One:} {\em Semiconductor constraints dictate new approaches.}
The ``free lunch'' of lower cost, higher performance transistors via
Dennard scaling \cite{end-dennard} and faster processors via Moore's Law
is slowing.
Moreover, the {\em de facto} assumption that integrating more devices onto a single
chip is always the best way to lower costs and maximize performance 
no longer holds.
Individual transistor costs are now flat to rising as feature sizes 
approach one nanometer, due to the interplay of chip yields on 300nm wafers and increasing fabrication facility costs.
Today, the investment needed to build state of the art facilities is denominated in billions of dollars per facility.

As recent geopolitical events have shown, there are substantial social, political, economic, and 
national security risks for any country or region that lacks a robust silicon fabrication ecosystem.
Fabless semiconductor firms rightly focus on design and innovation, but manufacturing those designs depends on reliable access to state of the art fabrication facilities, as the ongoing global semiconductor shortage has shown.
In the U.S., the the U.S. CHIPS Act and its successors, which would provide government support, are topics of intense political debate, with similar conversations underway in the European Union.
Finally, Intel, TSMC, and GlobalFoundries recently announced plans to build new chip fabrication facilities in the U.S., each for different reasons.

Optimization must balance chip fabrication facility costs, now
near \$10B at the leading edge, chip yield per wafer, and chip
performance.
This complex optimization process has rekindled interest in packaging
multiple chips, often fabricated with distinct processes and feature
sizes.
Such chiplets \cite{chiplet1,chiplet2} are more than a way to mix and match capabilities from multiple 
sources, they are an economic and engineering reaction to the interplay of chip defect rates, 
the cadence of feature size reductions, and semiconductor manufacturing costs.
However, this approach requires academic, government, and industry
collaborations to establish interoperability standards (e.g., the Open Domain-Specific 
Architecture (OSDA) project \cite{odsa} within the Open Compute Project \cite{opencompute}).
{\em Only with open chiplet standards can the best ideas from multiple sources be integrated effectively, in innovative ways, to develop next-generation HPC architectures.}

{\bf Maxim Two:}
{\em End-to-end hardware/software co-design is essential.}
Leveraging the commodity semiconductor ecosystem has led to an HPC
monoculture, dominated by x86-64 processors and GPU accelerators.
Given current semiconductor constraints, substantially increased system
performance will require more intentional end-to-end co-design \cite{birn}, from device physics to applications. 
Most recently, Japan's collaborative development of the Fugaku supercomputer, based on custom processors with ARM cores and vector instructions \cite{fugaku}, demonstrated the power of
application-driven end-to-end co-design.
Similar application driven co-designs were evident in the recent batch of AI hardware startup mentioned above as well as the cloud vendor accelerators. 
Such co-design means more than encouraging tweaks of existing products or product plans.  
Rather, it means looking holistically at the problem space, then envisioning, designing, testing, and fabricating appropriate solutions.
{\em In addition to deep partnership with hardware vendors and cloud ecosystem operators,
end-to-end co-design will require substantially expanded government investment in basic research and development, unconstrained by forced deployment timelines.}

Many of the most interesting innovations are happening in the world of AI hardware startups, cloud servers, and smartphones, where purpose-built chips and chiplets are being developed, based on detailed analysis of application demands.
One need look no further than AI hardware startup companies (e.g., Graphcore \cite{graphcore}, Groq \cite{groq}, and Cerebras \cite{cerebras}), cloud service vendor designs (e.g., Google TPUs \cite{tpu} and Amazon Gravitons), or 
smartphone SoCs (e.g., Apple's A15) for inspiration, and at the ferment in edge devices and AI.

{\bf Maxim Three:}
{\em Prototyping at scale is required to test new ideas.}
Semiconductors, chiplets, AI hardware, cloud innovations -- the computing system is now in great flux, and not for the first time.
As Figure  
\ref{fig:timeline} shows, the 1980s and 1990s were filled with innovative computing research projects and companies
that built novel hardware, new programming tools, and system software at large scale.
{\em To escape the current HPC monoculture and built systems better suited to current and emerging
scientific workloads, we must build real hardware and software prototypes at scale, not just
incremental ones, but ones that truly test new ideas using custom silicon and associated software.}
Implicitly, this means accepting the risk of failure, including at substantial scale, drawing insights 
from the failure, and then building lessons based on those insights.
A prototyping project that must succeed is not a research project; it is a product development.

Building such prototypes, whether in industry, national laboratories, or academia, depends on 
recruiting and sustaining integrated research teams –- chip designers, packaging engineers, system software developers, 
programming environment developers, and application domain experts –- in an integrated end-to-end way.
Such opportunities then make it intellectually attractive to work on science and engineering problems, particularly when there are industry partnerships and opportunities for translation
of research ideas into practice. 

{\bf Maxim Four:}
{\em The space of leading edge HPC applications is far broader now than in the past.}
Leading edge HPC originated in domains dominated by complex optimization
problems and solution of time-dependent partial differential equations on complex meshes.  
Those domains will always matter, but other areas of advanced computing 
in science and engineering are of high and growing importance.  
As an example, the {\em Science 2021 Breakthrough of the Year} \cite{breakthrough} was for
AI-enabled protein structure prediction \cite{alphafold}, with transformative implications for biology and biomedicine. 

Even in traditional HPC domains, the use of AI for data set reduction
and reconstruction and for PDE solver acceleration, is transforming
computational modeling and simulation.
The deep learning methods developed by the cloud  companies are changing the course of computational science, and university collaborations are growing. The University of Washington, with help from Microsoft Azure on protein-protein interaction \cite{proteinprotein}, is part of a bioscience revolution.  In other areas,  OpenAI is showing  that deep learning can solve challenging Math Olympiad problems \cite{math}. In astrophysics, deep learning is being used to classify galaxies \cite{galaxies}, generative adversarial networks (GANs) \cite{gan} have been used to  understand groundwater flow in superfund sites \cite{superfund},  and deep neural networks have been trained to help design non-photonic structures \cite{nanophotonic}.  This past year, the flagship conference of supercomputing (SC2021) had over 20 papers on neural networks in its highly selective program.
{\em The HPC ecosystem is expanding and engaging new domains and approaches in deep learning, creating new and common ground with cloud service providers.}

{\bf Maxim Five.} 
{\em Cloud economics have changed the supply chain ecosystem.}
The largest HPC systems are now dwarfed by the scale of commercial cloud infrastructure and social media company deployments.
A \$500M supercomputer acquisition every five years provides
limited financial leverage with chip vendors and system integrators relative to the billions of dollars spent each
year by cloud vendors.
Driven by market economics, computing hardware and software vendors, themselves small relative to the large cloud vendors,
now respond most directly to cloud vendor needs.

In turn, government investment (e.g., the United States Department of Energy 
(DOE) Exascale DesignForward, FastForward, and  PathForward programs
\cite{pathforward}, and the European Union's HPC-Europa3 \cite{eu}) 
are small compared to the scale of commercial cloud investments and their leverage with 
those same vendors. 
For example, HPC-Europa3, fully funded under the EU's Eighth Framework Programme, better known as Horizon 2020,
has a budget of only €9.2M \cite{eu}.
Similarly, the US DOE's multiyear investment of \$400M through the FastForward, DesignForward, 
and  PathForward programs as part of the Exascale Computing 
Project (ECP) targeted innovations in reduced power consumption, resilience, improved network and system integration. The DOE only supplied approximately \$100M in NRE for each of the  exascale systems under construction, while the cloud companies invested billions.
Market research  \cite{hyperion} suggests that China, Japan, the United States, and the European Union may each procure 1-2 exascale class systems per year, each estimated at approximately \$400M.

The financial implications are clear.
The government and academic HPC communities have limited leverage
and cannot influence vendors in the same ways they did in the past. 
{\em New, collaborative models of partnership and funding are needed that recognize and
embrace ecosystem changes and their implications, both in use of cloud services and collaborative
development of new system architectures.}
As we have emphasized, the smartphone and cloud services vendors now dominate the computing ecosystem.
We realize this may be heretical to some, but there are times and places where commercial cloud services
can be the best option to support scientific and engineering computing needs.
The performance gaps between cloud services and HPC gaps have lessened substantially over the past decade,
as shown by a recent comparative analysis \cite{berkeley}.
Moreover, HPC as a service is now both real and effective, both because
of its performance and the rich and rapidly expanding set of hardware capabilities and software services.
The latter is especially important; cloud
services offer some features not readily available in the HPC software ecosystem.

%
Some in academia and national laboratory community will immediately say, 
''But, we can do it cheaper, and our systems are bigger!''
Perhaps, but those may not be the appropriate perspectives.
Proving such claims means being dispassionate about technological 
innovation, NRE investments, and opportunity costs.
In turn, this requires a mix of economic and cultural realism in making 
build versus use decisions and taking an expansive view of the application
space, unique hardware capabilities, and software tools.
Opportunity costs are real, though not often quantified in academia or government.
Today, capacity computing (i.e., solving an ensemble of smaller problems) can easily be 
satisfied with a cloud-based solution, and
on-demand, episodic computing of both capacity and large-scale scientific computing
can benefit from cloud scaling.

{\bf Maxim Six:}
{\em The societal implications of technical issues really matter.}
Large-scale computational models and AI-driven data analysis 
now pervade all aspects of business, e-commerce, and science and 
engineering, with concomitant societal implications.
Even more importantly, how countries and regions respond to these
issues, from semiconductor fabrication through appropriate use, is
likely to shape their competitive futures.
Simply put, HPC systems and their diverse applications now underpin everything
from weather forecasts through building and infrastructure design to drug
discovery.  As designers, operators, and users of HPC systems,
we must both acknowledge and engage actively in the ethics of their use.
%

\section{Conclusions}
\label{sec:conclusions}
The computing ecosystem is in enormous flux, creating both opportunities and challenges for the future of advanced computing.
It seems increasingly unlikely that future high-end HPC systems will be procured and assembled solely by commercial integrators 
from commodity components.  
Rather, future advances will require embracing end-to-end design, testing and evaluating advanced prototypes,
and partnering strategically.
These are likely to involve (a) collaborative partnerships among academia, government laboratories, chip vendors, and cloud providers, (b) increasingly 
bespoke systems, designed and built collaboratively to support key scientific and engineering workload needs, 
or (c) a combination of these two.

Put another way, in contrast to midrange systems, leading edge, HPC  computing systems are increasingly 
similar to large-scale scientific instruments (e.g., Vera Rubin Observatory, the LIGO gravity wave detector,  Large Hadron Collider, or the Square Kilometer Array),
with limited economic incentives for commercial development.
Each contains commercially designed and constructed technology, but they also contain 
large numbers of custom elements for which there is no sustainable business model.  
Instead, we build these instruments because we want them to explore open scientific questions, 
and we recognize that their design and construction requires both government investment and innovative private
sector partnerships. 

Investing in the future is never easy, but it is critical if we are to continue to develop and deploy new generations of high-performance computing systems, ones that leverage economic shifts, commercial practice, and emerging technologies.
Let us be clear.  The price of innovation keeps rising, the talent is following the money, and many of the traditional players – companies and countries – are struggling to keep up.  
Intel co-founder, Andrew Grove was only partially right when he said,
\begin{quote}
\textit{
    Success breeds complacency.  Complacency breeds failure.  
    Only the paranoid survive.
    }
\end{quote}
It is true that only the paranoid survive, but the successful competitors also need substantial financial resources and a talented workforce, both aligned with market forces and technological opportunities.

\section*{Acknowledgments}
The authors are grateful for thoughtful feedback on earlier drafts of this article from Doug Burger (Microsoft), Tony Hey (UK Science and Technology Facilities Council), and John Shalf (Lawrence Berkeley National Laboratory).

\bibliographystyle{IEEEtran}
\bibliography{references}

\end{document}